# Dimensionality-Driven Electronic and Orbital Transitions Mediating Interfacial Magnetism in LaNiO$_3$/CaMnO$_3$ Observed *In Situ*

B-A. Courchene[1], A. Hampel[2], S. Beck[2], J. R. Paudel[1,3], J. D. Grassi[1,4], L. A. Lapinski[1], A. M. Derrico[1,5], M. Terilli[6], M. Kareev[6], C. Klewe[7], A. Gloskovskii[8], C. Schlueter[8], S. K. Chaluvadi[9], F. Mazzola[9,10], I. Vobornik[9], P. Orgiani[9,11], J. Chakhalian[2], A. J. Millis[2,12], and A. X. Gray[1,*]

[1] *Department of Physics, Temple University, Philadelphia, Pennsylvania 19122, USA*
[2] *Center for Computational Quantum Physics, Flatiron Institute, New York, New York 10010, USA*
[3] *Chemical Sciences Division, Lawrence Berkeley National Laboratory, Berkeley, California 94720, USA*
[4] *Department of Physics, Cornell University, Ithaca, New York 14853, USA*
[5] *Department of Physics, University of California, Berkeley, California 94720, USA*
[6] *Department of Physics and Astronomy, Rutgers University, Piscataway, New Jersey 08854, USA*
[7] *Advanced Light Source, Lawrence Berkeley National Laboratory, Berkeley, California 94720, USA*
[8] *Deutsches Elektronen-Synchrotron, DESY, 22607 Hamburg, Germany*
[9] *CNR-IOM Istituto Officina dei Materiali, Trieste 34149, Italy*
[10] *Dipartimento di Fisica e Astronomia, Università degli Studi di Padova, Padua 35131, Italy*
[11] *AREA Science Park, Trieste 34149, Italy*
[12] *Department of Physics, Columbia University, New York, New York 10027, USA*
*email: axgray@temple.edu

Emergent magnetic states at oxide interfaces arise from the interplay of charge transfer, orbital reconstruction, and dimensional confinement, offering a route to engineered correlated-electron behavior in nanoscale spintronic materials. Here, we combine in situ synthesis, polarization-dependent angle-resolved photoelectron spectroscopy, X-ray magnetic circular dichroism, and first-principles electronic-structure calculations to investigate LaNiO$_3$/CaMnO$_3$ superlattices. We show that reducing the LaNiO$_3$ thickness drives a metal-insulator transition accompanied by loss of electronic coherence and an orbital-polarization crossover in the ultrathin limit. These changes weaken charge transfer across the interface and suppress the interfacial Mn magnetic moment in CaMnO$_3$, revealing that the emergent ferromagnetic state is directly governed by electronic confinement in LaNiO$_3$. The insulating state and orbital reconstruction are reproduced by density functional theory combined with dynamical mean-field theory. Together, these results establish a direct and tunable coupling among electronic, orbital, and magnetic degrees of freedom in oxide heterostructures.

**Introduction**

Interfaces between correlated oxides provide a powerful platform where charge, orbital, and lattice degrees of freedom can interact to produce electronic and magnetic states not realized in the bulk [1-5]. Advances in atomic-layer synthesis now allow these interfacial behaviors to be engineered through dimensional confinement, epitaxial strain, and controlled defect incorporation, offering a pathway to design quantum materials with tailored functionalities [6-11]. These systems therefore provide an opportunity to investigate how electronic, orbital, and magnetic degrees of freedom become intertwined at the atomic scale, a central theme in the study of correlated oxide heterostructures [12-14].

LaNiO$_3$/CaMnO$_3$ (LNO/CMO) superlattices exemplify this principle: although neither LNO nor CMO is ferromagnetic in bulk, their interface hosts a robust ferromagnetic ground state [15]. This magnetism arises through charge transfer from itinerant Ni $3d$ $e_g$ states in LNO to Mn cations in CMO, stabilizing Mn$^{3+}$-Mn$^{4+}$ double-exchange interactions [16]. Because this exchange mechanism relies on metallic LNO, the thickness-driven metal-insulator transition (MIT) in ultrathin LNO layers provides an intrinsic means of tuning the interfacial magnetic state [17-19]. The MIT in LNO is itself of broad interest in correlated-electron physics, as it reflects the interplay of bandwidth reduction, crystal-field effects, and many-body interactions that emerge under reduced dimensionality [20-22].

Although epitaxial LNO single films grown on crystalline substrates exhibit critical thicknesses for the MIT ranging from 1.5 to 4 unit cells (u.c.) [22-24], the corresponding evolution of LNO electronic and orbital structure within the LNO/CMO heterostructure remains unresolved. Prior angle-resolved photoelectron spectroscopy (ARPES) studies of epitaxial LNO single films have shown that ultrathin LNO loses quasiparticle coherence at 2 u.c., displays quasi-one-dimensional spectral features under tensile strain below 3-4 u.c., and develops a reduced mean free path with strongly suppressed coherent spectral weight below 5 u.c. [22,23,25]. These observations highlight the strong sensitivity of the LNO electronic structure to reduced dimensionality, strain, and growth conditions. Because heterostructuring can additionally modify crystal symmetry, octahedral rotations, and magnetic exchange pathways, characterizing the internal electronic and orbital structure of LNO within LNO/CMO is essential for understanding and controlling its interfacial magnetic behavior. Resolving this evolution requires not only direct spectroscopic measurements but also a theoretical framework that captures the role of strong correlations in the ultrathin limit. Addressing this challenge also demands state-of-the-art experimental capabilities, since ARPES measurements must be performed on superlattices grown *in situ* directly at a synchrotron beamline, where photon-energy tuning enables controlled access to $k_z$ but imposes far more stringent constraints than typical *in situ* studies on single films.

Here, by combining *in situ* pulsed-laser deposition (PLD) epitaxy with polarization-dependent synchrotron-based ARPES, together with complementary element-specific X-ray magnetic circular dichroism (XMCD) measurements, we determine the critical thickness for the LNO MIT in the heterostructure and track the associated evolution of orbital polarization and magnetic response. In the ultrathin limit, LNO exhibits a fully gapped insulating state and an orbital polarization crossover characterized by enhanced in-plane $d_{x2-y2}$ occupancy within the 3$d$ $e_g$ manifold. Both features are reproduced by first-principles electronic-structure calculations performed using density functional theory combined with dynamical mean-field theory (DFT+DMFT), which provide a consistent theoretical framework for the dimensionality-driven evolution of electronic properties in LNO. Together, these results demonstrate how reduced dimensionality in LNO mediates coupled electronic, orbital, and magnetic reconstruction at the LNO/CMO interface, with broader implications for designing emergent magnetic states in correlated oxide heterostructures.

## Results

Figure 1a shows a schematic diagram of the epitaxial LNO/CMO superlattices investigated in this study. Alternating layers of LNO (blue) and CMO (purple) were grown on LaAlO$_3$ (001) substrates using *in situ* PLD [26]. Four superlattices were synthesized, with LNO layer thicknesses of 6, 4, 3, or 1 u.c. and CMO layers of 3 u.c., except in the 6-u.c.-LNO superlattice (first sample), in which the CMO layer thickness was found to be slightly higher, at 4 u.c. This slight variation does not affect the conclusions presented in this study. All samples terminate in LNO to facilitate direct ARPES measurements (see Methods for further growth details).

In addition to *in situ* ARPES, the structural and chemical quality of the superlattices was verified through a comprehensive set of *in situ* and *ex situ* characterization techniques. Low-energy electron diffraction (LEED) patterns acquired *in-situ* immediately after growth, shown in Figure S1a of the Supplementary Information, confirmed the epitaxial order and high surface crystallinity of the topmost LNO layer. Scanning Tunneling Microscopy (STM) *in-situ* imaging (Fig. S1b) showed smooth morphology consistent with layer-by-layer growth with RMS roughness of 0.3 nm (~ 1 u.c.). High-resolution X-ray diffraction (XRD) *ex-situ* measurements (Figs. S2a-d) revealed well-defined superlattice satellite peaks and thickness fringes, confirming coherent epitaxy, the designed superlattice periodicity, and the overall crystalline quality of all four samples. High-resolution X-ray reflectivity (XRR) measurements (Figs. S2e-h) and best fits to the data confirmed high quality of the samples and the target thicknesses of all layers. Finally, *ex-situ* bulk-sensitive hard X-ray photoelectron spectroscopy (HAXPES) [27] (Fig. S3) verified the expected chemical composition of the LNO/CMO superlattices. Further details of the HAXPES measurements are provided in the Methods section. Together, these characterization results (see Supplementary Information) establish the structural integrity, interface sharpness, and chemical composition of the samples used in this study.

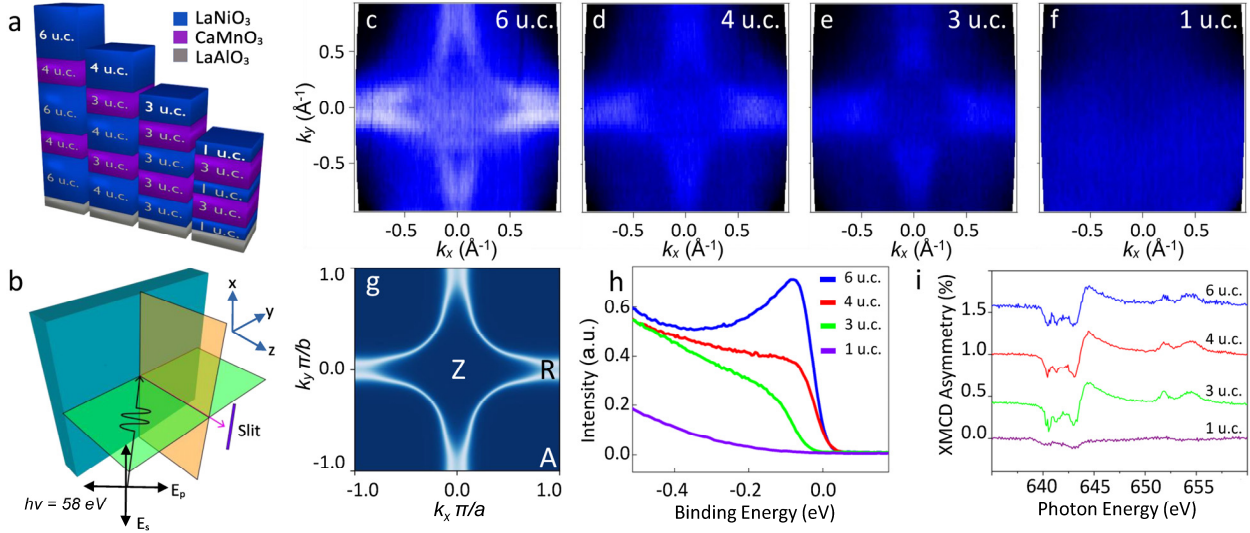

**Fig. 1 | Dimensionality-driven evolution of electronic structure and interfacial magnetism in LNO/CMO superlattices. a,** Schematic of the epitaxial LNO/CMO superlattices. Alternating LNO (blue) and CMO (purple) layers are grown on LaAlO$_3$ (001), with the CMO thickness fixed at 3-4 u.c. and the LNO thickness varied; all superlattices are terminated with an LNO layer. **b,** ARPES measurement geometry. The photon beam impinges on the surface-terminating LNO layer at 45° using linearly polarized light with E$_s$ (electric field fully in-plane) and E$_p$ (electric field with both in-plane and out-of-plane components). **c-f,** ARPES Fermi-surface intensity maps at the Z point ($h\nu$ = 58 eV, T = 77 K) for LNO thicknesses of 6, 4, 3, and 1 u.c., respectively. Hole-like pockets around the R points, characteristic of metallic LNO, are clearly visible for 6, 4, and 3 u.c., but are absent at 1 u.c., where no Fermi-surface features are observed. **g,** DFT+DMFT-calculated Fermi-surface spectral weight of LNO in the ZRA plane. The calculation reproduces the topology and momentum-dependent spectral weight observed experimentally for the 6 u.c. superlattice. **h,** Energy-distribution curves (EDCs) for all four LNO thicknesses. A sharp quasiparticle peak at the Fermi level for 6 u.c. is progressively suppressed with decreasing LNO thickness and vanishes at 1 u.c. **i,** XMCD asymmetry at the Mn $L$ edge as a function of photon energy for the same series of superlattices (T = 20 K). A strong XMCD signal is observed for 6, 4, and 3 u.c., whereas the XMCD asymmetry is strongly suppressed at 1 u.c.

ARPES experiments were conducted at the APE-LE beamline of the Elettra synchrotron [28], where the superlattices were grown and transferred under ultrahigh-vacuum conditions to avoid surface contamination. The experimental geometry is shown in Figure 1b, with the photon beam impinging on the topmost LNO layer at an incidence angle of 45°. Spectra were acquired using linearly polarized light ($E_p$ and $E_s$) at photon energies corresponding to the Γ and Z planes in $k_z$, with the Z-point measurements ($hv = 58$ eV) forming the basis of the Fermi-surface analysis discussed below. Because ARPES directly probes the electronic states of the surface-terminating LNO layer, which remains part of the LNO/CMO stack, these measurements provide a layer-resolved view of how heterostructure-induced electronic reconstruction manifests in the topmost LNO layer. Additional experimental details are provided in the Methods section.

Figures 1c through 1f present the Fermi-surface intensity maps obtained at the Z point in $k_z$. In the 6 u.c. LNO/CMO superlattice (Figure 1c), the characteristic hole pockets centered at the R points, expected for the LNO Fermi surface at $k_z = \pi/c$, are clearly resolved. These features remain visible in the 4 u.c. and 3 u.c. films (Figures 1d and 1e), although with reduced spectral intensity as the LNO layer thickness is decreased. In contrast, the 1 u.c. superlattice (Figure 1f) exhibits no discernible Fermi-surface features, indicating a complete suppression of the coherent Ni $3d$ $e_g$ quasiparticle. This collapse of quasiparticle spectral weight at the Fermi level corresponds to the onset of an insulating state in ultrathin LNO. The observed thickness-dependent suppression of Ni $3d$ $e_g$ coherence is consistent with a reduction in bandwidth and an increase in correlation strength as LNO approaches the two-dimensional limit, a trend that is captured qualitatively within the DFT+DMFT framework [29]. The observation of a 3 u.c. critical thickness within the heterostructure is consistent with the canonical thickness-driven MIT reported for epitaxial LNO single films on crystalline substrates [22-25], indicating that the dimensionality-driven coherence collapse is robust even in the presence of interfacial coupling to CMO and the associated coherent epitaxial strain.

To compare the experimental data with theoretical expectations, Figure 1g shows a DFT+DMFT calculation of the Fermi surface of LNO evaluated in the ZRA plane under compressive strain (see Methods for computational details). The calculation reproduces both the Fermi-surface topology and the momentum-dependent spectral-weight distribution observed in the 6 u.c. film, supporting the assignment of the experimental features to Ni $3d$ $e_g$ orbital character. The good agreement between theory and experiment further indicates that the ferromagnetism is interfacial and does not measurably reconstruct the LNO-derived band structure.

Experimental energy-distribution curves (EDCs) extracted at the $k_F$ locations of these features are shown in Figure 1h. The 6 u.c. film displays a sharp quasiparticle peak at E = 0 eV, characteristic of a coherent metallic state. This peak becomes progressively suppressed as the LNO thickness is reduced and shifts slightly away from $E_b$ = 0 at 3 u.c., indicating the onset of the dimensionality-driven MIT at 3 u.c., before disappearing entirely at 1 u.c., where the fully gapped insulating state is reached [22,25,29]. This trend is corroborated by complementary bulk-sensitive valence-band HAXPES measurements probing the full depth of the superlattice. As shown in Supporting Figure S4, *ex situ* HAXPES likewise reveals a systematic suppression of the strongly hybridized near-Fermi-level Ni $3d$ $e_g$ and $t_{2g}$ states with decreasing LNO thickness. The complete loss of coherent quasiparticle weight at 1 u.c. is also consistent with the orbital-polarization crossover discussed in the following section, linking electronic coherence and orbital occupancy in the ultrathin limit.

Figure 1i shows the XMCD asymmetry at the Mn $L_{2,3}$ edge as a function of photon energy. XMCD measurements were performed at the high-resolution Magnetic Spectroscopy and Scattering beamline 4.0.2 of the Advanced Light Source [30] in the bulk-sensitive luminescence-yield detection mode at T = 20 K (see Methods for further experimental details). For the 6 u.c., 4 u.c., and 3 u.c. LNO/CMO superlattices, where LNO remains metallic and capable of supplying itinerant charge carriers, pronounced XMCD asymmetry is observed near 645 eV (Mn $L_3$ edge), indicating a finite interfacial Mn magnetic moment and ferromagnetic alignment. The magnitude of the XMCD signal increases slightly but systematically as the LNO layer is thinned from 6 u.c. to 3 u.c., consistent with enhanced charge transfer into the interfacial CMO layer as dimensional confinement reduces the bandwidth of LNO [29,31]. In the 1 u.c. superlattice, however, the XMCD signal is almost completely suppressed, which demonstrates that the interfacial Mn moment collapses once LNO becomes strongly insulating and is no longer able to supply charge to CMO to support $Mn^{3+}$ to $Mn^{4+}$ double-exchange interactions.

A small residual XMCD asymmetry remains detectable in the 1 u.c. superlattice. This remnant magnetic signal is consistent with prior observations in similar LNO/CMO heterostructures, where a weak interfacial moment of order 0.3 μB per Mn has been attributed to $Ni^{2+}$-$Mn^{4+}$ superexchange interactions enabled by oxygen vacancies that migrate toward the LNO/CMO interface under polar mismatch [16]. Such defect-mediated contributions, together with possible minor chemical intermixing, represent secondary effects that do not alter the overall suppression of the double-exchange-driven ferromagnetism in the insulating regime, and they can be minimized through optimized growth conditions and strain control.

Figure 2a and 2b show DMFT-calculated spectral functions for bulk-like and monolayer LNO, respectively, highlighting the strong influence of reduced dimensionality on the electronic structure. The bulk-like calculation (Figure 2a) exhibits finite spectral weight at the Fermi level and well-defined Ni $3d$ $e_g$-derived bands, consistent with the metallic ground state observed in thick LNO films and in the 6 u.c. and 4 u.c. LNO/CMO superlattices. In contrast, the monolayer calculation (Figure 2b) shows a fully developed gap at the Fermi level together with the complete suppression of coherent quasiparticle spectral weight, indicating an insulating ground state in the ultrathin limit. The monolayer calculation incorporates the reduced coordination environment and modified crystal-field splitting characteristic of the ultrathin limit, enabling a direct comparison with the experiment. In addition to reproducing the insulating gap, the calculation captures the redistribution of spectral weight away from the Fermi level, matching the evolution observed experimentally in the ARPES maps. The emergence of this gapped electronic structure in the monolayer calculation mirrors the ARPES response of the 1 u.c. LNO/CMO superlattice, demonstrating that first-principles DFT+DMFT captures both the collapse of coherence and the onset of insulating behavior as the LNO layer is reduced to a single unit cell.

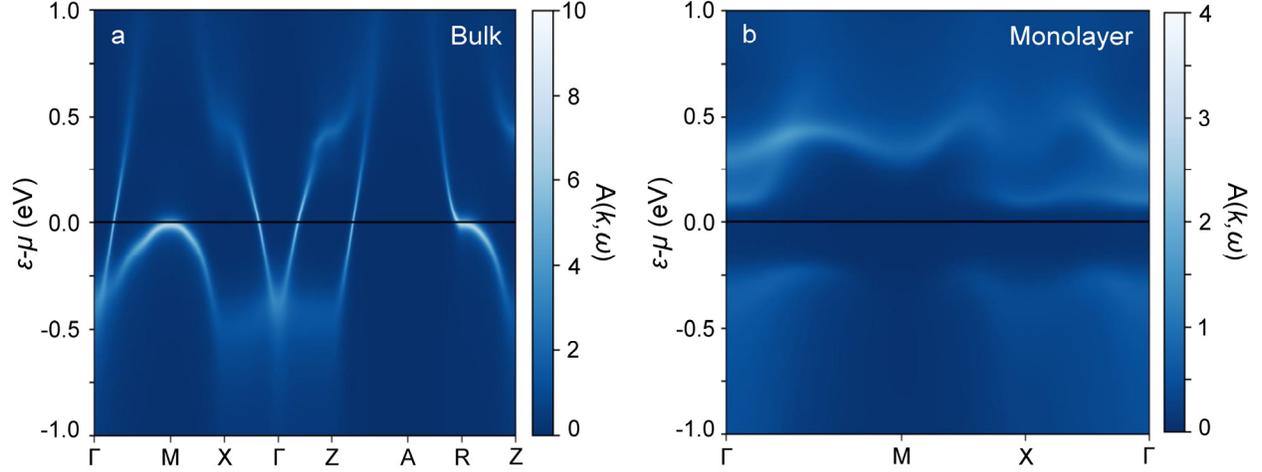

**Fig. 2 | Dimensionality-driven evolution of the LNO spectral function from DFT+DMFT. a,** Unfolded $k$-resolved spectral function $A(k,\omega)$ of bulk-like LNO calculated using DFT+DMFT along a high-symmetry path in the Brillouin zone. Finite spectral weight at the Fermi level (black horizontal line) and well-defined Ni $3d$ $e_g$-derived bands are consistent with a correlated metallic state. **b,** Corresponding DFT+DMFT spectral function for monolayer LNO evaluated along the same high-symmetry $k$-path. The opening of a gap signifies an insulating state in the ultrathin limit, in agreement with the ARPES observations for the 1 u.c. LNO/CMO superlattice.

Together, the ARPES, DFT+DMFT, and XMCD measurements establish that coherent Ni 3$d$ $e_g$ quasiparticle, metallic electronic states in LNO, and interfacial ferromagnetism in LNO/CMO superlattices persist down to an LNO thickness of 3 u.c., but vanish upon reaching 1 u.c. The fully gapped insulating state observed experimentally in the ultrathin (1 u.c.) limit is quantitatively reproduced by the DMFT spectral functions, which capture both the loss of quasiparticle coherence and the opening of an energy gap as the system approaches the monolayer limit. This combined experimental and theoretical picture identifies a critical thickness of 3 u.c. for the MIT of LNO within the heterostructure and demonstrates that dimensional confinement in LNO directly mediates the coupled electronic, orbital, and magnetic reconstruction at the LNO/CMO interface. These results establish the electronic and magnetic boundary conditions that govern the behavior of ultrathin LNO within oxide heterostructures and provide the foundation for understanding the orbital polarization crossover and dimensional tuning mechanisms examined in the following sections.

To investigate how the Ni 3$d$ $e_g$ orbital character evolves under dimensional confinement in LNO/CMO superlattices, we performed polarization-dependent ARPES linear dichroism measurements as a function of LNO thickness. In Figures 3a through 3d, we present the momentum-resolved ARPES linear dichroism spectra for all four LNO thicknesses, starting with the 6 u.c. superlattice in Figure 3a and moving from left to right in descending order down to the 1 u.c. sample in Figure 3d. For each thickness, the dichroic signal is obtained by subtracting the normalized $I_{Es}$ spectrum from the normalized $I_{Ep}$ spectrum. In our experimental geometry, the linear vertical ($E_s$) polarization lies fully in the sample plane and is odd with respect to the mirror plane within the scattering plane, so it predominantly couples to in-plane components of the Ni 3$d$ $e_g$ states. The linear horizontal ($E_p$) polarization impinges on the sample at 45º and has both in-plane and out-of-plane components, so it samples orbital states with both in-plane and out-of-plane symmetry. As a result, the $I_{Ep} - I_{Es}$ linear dichroism provides a qualitative measure of the additional contribution from out-of-plane Ni $3d$ $d_{z2}$ orbital character relative to the predominantly in-plane response emphasized by $E_s$.

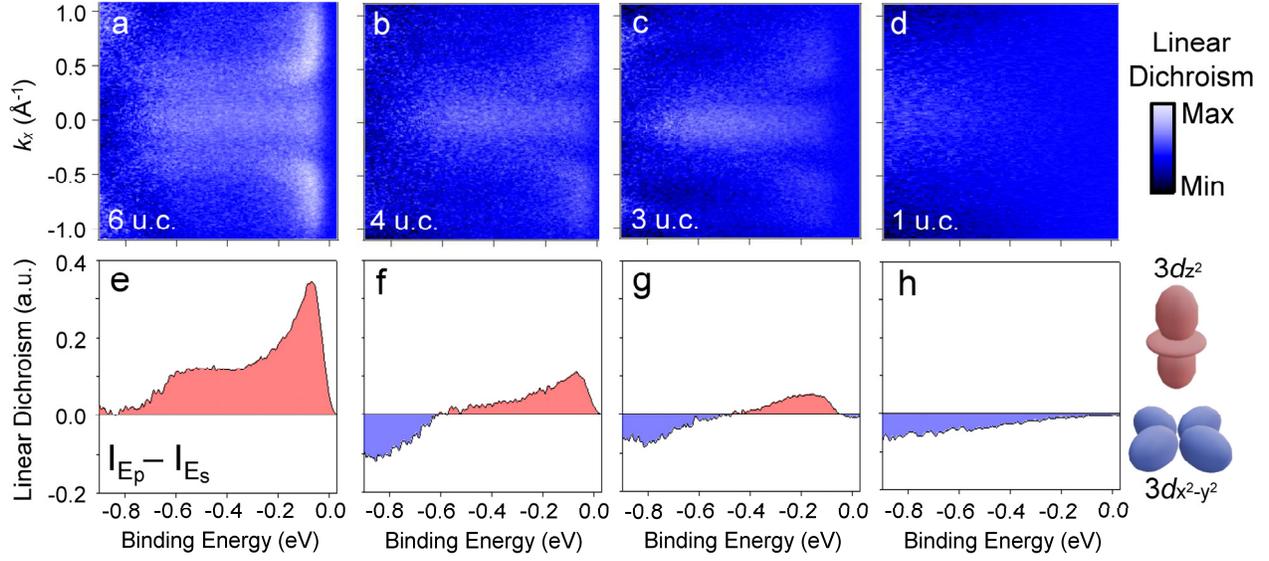

**Fig. 3 | Polarization-dependent ARPES linear dichroism in LNO/CMO superlattices. a-d,** Momentum-resolved ARPES linear dichroism maps $I_{Ep} - I_{Es}$ at the Z point for LNO thicknesses of 6, 4, 3, and 1 u.c., respectively. For each sample, the dichroic signal is obtained by subtracting the normalized $I_{Es}$ spectrum (electric field fully in-plane) from the normalized $I_{Ep}$ spectrum (electric field with both in-plane and out-of-plane components), so that positive intensity highlights the additional contribution from out-of-plane Ni $3d$ $d_{z2}$-derived states relative to the predominantly in-plane response emphasized by $E_s$. **e-h,** Corresponding angle-integrated $I_{Ep} - I_{Es}$ linear dichroism spectra for the same set of LNO thicknesses, obtained by integrating over the momentum window used in panels a-d. Positive (red) and negative (blue) intensities indicate the relative dominance of $E_p$- and $E_s$-emphasized components, respectively.

At 6 u.c. (Figure 3a), the momentum-resolved linear dichroism shows strong positive intensity near the Fermi level, indicating a substantial additional contribution from orbitals with out-of-plane character to the low-energy electronic structure. As the LNO thickness is reduced to 4 u.c. and 3 u.c. (Figures 3b and 3c), the overall dichroic intensity decreases, reflecting a systematic modification of the Ni $3d$ $e_g$ orbital composition with increasing confinement. In the ultrathin 1 u.c. superlattice (Figure 3d), the dichroic signal is strongly suppressed and nearly disappears across the entire measured momentum range. This progressive reduction and eventual near-vanishing of the momentum-resolved dichroism with decreasing thickness points to a substantial reconstruction of Ni 3$d$ $e_g$ orbital contributions in the vicinity of the Fermi level as LNO approaches the ultrathin insulating regime.

Figures 3e through 3h show the corresponding angle-integrated linear dichroism spectra for the same set of LNO thicknesses. These spectra are obtained by integrating the $I_{Ep}$ and $I_{Es}$ intensities over the same momentum window used for Figures 3a through 3d and then subtracting $I_{Es}$ from $I_{Ep}$. For the 6 u.c. superlattice (Figure 3e), the dichroism is positive over the entire energy range considered, which indicates that the components emphasized by $E_p$, including the out-of-plane Ni 3$d$ $d_{z2}$ contribution, dominate the low-energy spectral weight relative to the response emphasized by $E_s$. As the LNO thickness is reduced, the dichroic spectra evolve systematically. The positive area near the Fermi level is reduced, the curves cross through zero, and by 1 u.c. (Figure 3h) the area under the curve becomes entirely negative. This evolution reflects an increasing relative weight of the $E_s$-emphasized in-plane components and a redistribution of spectral weight in the $E_p$ response as the LNO layer becomes thinner. In particular, the reduction and eventual sign reversal of the dichroic signal near zero energy indicate that the additional contribution from out-of-plane Ni 3$d$ $d_{z2}$ orbital character to the $I_{Ep}$ spectra is strongly diminished as the LNO thickness decreases.

The experimental dichroism trends can be compared directly with DFT+DMFT calculations of the orbital-resolved spectral functions for bulk and monolayer orthorhombic LNO under compressive strain, shown in Figure 4. In these calculations, the dichroism $\Delta A (k,\omega)$ is defined as the difference between the total spectral function, $A_{tot} (k,\omega)$, and the spectral function projected onto the $d_{x2-y2}$ orbital, $A_{dx2-y2} (k,\omega)$. This quantity, $\Delta A (k,\omega) = A_{tot} (k,\omega) - A_{dx2-y2} (k,\omega)$, therefore isolates the contribution from the $d_{z2}$-derived states to the total spectral weight. This can be compared to the experimental observable $I_{Ep} - I_{Es}$ in which the out-of-plane orbital component is enhanced. Band-structure calculations indicate that, in the relevant Z-R momentum region, the states just below the Fermi level are dominated by $d_{x2\text{-}y2}$ character, with the out-of-plane $d_{z2}$ contribution increasing as the band approaches the R point. Figures 4a and 4b present the momentum-resolved dichroism obtained by DFT+DMFT for bulk-like and monolayer LNO, respectively, while Figure 4c shows the corresponding k-integrated dichroism for these two systems.

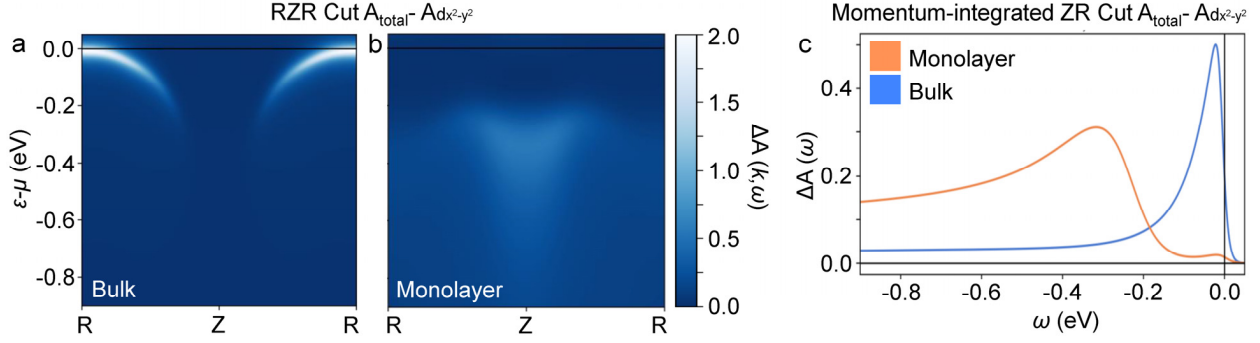

**Figure 4 | Theoretical orbital dichroism from DFT+DMFT for bulk-like and monolayer LNO. a, b,** Momentum-resolved dichroism $\Delta A\,(k,\omega) = A_{\text{tot}}\,(k,\omega) - A_{d_{x2\text{-}y2}}\,(k,\omega)$ calculated using DFT+DMFT for bulk-like (a) and monolayer (b) LNO under compressive strain. The dichroism is obtained by subtracting the spectral function projected onto the $d_{x2\text{-}y2}$ orbital from the total spectral function $A_{\text{tot}}\,(k,\omega)$, so that the resulting intensity isolates the contribution from $d_{z2}$-derived states. **c,** k-integrated dichroism $\Delta A\,(\omega) = A_{\text{tot}}\,(\omega) - A_{d_{x2\text{-}y2}}\,(\omega)$ for bulk-like (blue) and monolayer (orange) LNO, obtained by integrating along the same high-symmetry path as in **a** and **b**. The bulk-like case shows substantial $d_{z2}$-derived spectral weight near the Fermi level, whereas in the monolayer this weight is strongly suppressed at low energies and shifted to higher binding energies, mirroring the trend observed in the experimental ARPES linear dichroism.

For bulk-like LNO, the DMFT dichroism exhibits substantial spectral weight near the Fermi level, indicating a significant $d_{z2}$ contribution to the low-energy electronic structure. This behavior is consistent with the strong positive experimental dichroism observed for the 6 u.c. superlattice in Figures 3a and 3e. In the monolayer calculation, however, the dichroism shows a marked redistribution of spectral weight: the intensity near the Fermi level is strongly reduced, while a pronounced feature appears around -0.4 eV in Figure 4b. This indicates that the $d_{z2}$-derived spectral weight is shifted away from the Fermi level toward higher binding energies as the system approaches the ultrathin limit. Comparing the *k*-integrated dichroism in Figure 4c with the experimental angle-integrated dichroism in Figures 3e and 3h reveals the same qualitative trend. In the bulk-like case, the dichroism is dominated by states near zero energy, whereas in the ultrathin limit the near-Fermi-level dichroism is strongly suppressed and spectral weight at higher binding energy becomes comparatively more important.

The theoretically calculated dichroism provides a useful proxy for the experimental ARPES dichroism. In the experiment, the dichroic signal is defined as $I_{Ep} - I_{Es}$, whereas in the calculation it is defined as $A_{tot}(k,\omega) - A_{dx2-y2}(k,\omega)$. The $E_p$ spectrum does not represent the full total spectral weight, and the $E_s$ spectrum contains some residual sensitivity to both $d_{x2-y2}$ and $d_{z2}$ components. Nevertheless, because the experimental geometry is chosen so that $I_{Es}$ predominantly emphasizes in-plane orbital components, while $I_{Ep}$ contains both in-plane and out-of-plane contributions, the $I_{Ep} - I_{Es}$ dichroism serves as a reasonable measure for the additional $d_{z2}$-derived spectral weight that can be qualitatively compared to $A_{tot}(k,\omega) - A_{dx2-y2}(k,\omega)$.

The overall consistency between the bulk-like experiment and theory supports the interpretation that the low-energy spectral weight near the Fermi level is dominated by a coherent $d_{x2-y2}$ quasiparticle resonance in the bulk. The common experimental trend of suppressed near-Fermi-level dichroism and enhanced relative weight at higher binding energies in the ultrathin limit then points to a collapse of this coherent low-energy weight as the LNO layer thickness decreases, a picture consistent with the Mott-insulating-like suppression of the quasiparticle resonance seen in the DFT+DMFT calculations for the monolayer.

Taken together, the polarization-dependent ARPES linear dichroism and the DFT+DMFT orbital-resolved spectral functions demonstrate a dimensionality-driven orbital polarization crossover in LNO. In the bulk-like regime, both $d_{x2-y2}$ and $d_{z2}$ orbitals contribute to the near-Fermi-level spectral weight, whereas in the ultrathin limit the $d_{z2}$ contribution near the Fermi level is strongly suppressed and its spectral weight is shifted to higher binding energies. This evolution is consistent with a modified hybridization environment expected under strong confinement, and it provides a microscopic orbital perspective that complements the coherence collapse and gap opening identified in the previous section.

## Discussion

We combine *in situ* polarization-dependent ARPES, XMCD, and first-principles DFT+DMFT calculations to determine how dimensional confinement in LNO/CMO superlattices governs the electronic and magnetic properties at the interface. We find that LNO undergoes an onset of thickness-driven MIT at 3 u.c., with complete collapse of quasiparticle coherence and Fermi-level spectral weight at 1 u.c. This electronic reconstruction is accompanied by a suppression of the interfacial Mn moment, establishing a direct correspondence between the metallicity of LNO and the double-exchange-driven interfacial ferromagnetism in CMO. Polarization-dependent ARPES further reveals a dimensionality-driven orbital polarization crossover, in which the Ni $3d$ $d_{z2}$ contribution to the low-energy spectral weight is strongly reduced in the ultrathin limit. First-principles DFT+DMFT calculations reproduce both the insulating gap and the redistribution of orbital spectral weight, providing a consistent microscopic picture of the confinement-induced evolution of charge and orbital degrees of freedom. These results identify the critical thickness governing interfacial magnetism in LNO/CMO and establish a quantitative framework for understanding dimensionality-controlled electronic phases in correlated oxide heterostructures. The ability to tune orbital, electronic, and magnetic interactions through confinement offers a promising route toward engineering correlated-oxide interfaces for future quantum and spintronic functionalities.

## Methods

**Film Growth and Sample Preparation.** Epitaxial LNO/CMO superlattices were synthesized using in situ PLD at the APE-LE beamline of the Elettra synchrotron [26]. Alternating layers of LNO and CMO were deposited on single-crystalline $LaAlO_3$ (001) substrates using a KrF excimer laser ($\lambda$ = 248 nm). The substrate temperature during growth was maintained at 720 °C, with an oxygen partial pressure of $1\times10^{-2}$ mbar. The laser repetition rate was 1 Hz, the substrate–target distance was 5 cm, and the laser fluence was maintained between 1.4 and 1.7 J cm$^{-2}$. Four superlattices were prepared, each consisting of 3-4 unit cells (u.c.) of CMO and LNO layers of thickness 6, 4, 3, and 1 u.c., with all samples terminating in LNO to enable direct ARPES measurements of the nickelate layer. Following growth, the samples were cooled to room temperature under the same oxygen background pressure. The long-range surface crystallinity and epitaxial order of the as-grown films were monitored in situ by low-energy electron diffraction (LEED) at 70 eV, confirming sharp diffraction features from the surface LNO layer. Growth rates of 0.10 Å/pulse for LNO and 0.08 Å/pulse for CMO were calibrated by X-ray reflectivity.

**ARPES Measurements.** ARPES measurements were carried out at the APE-LE beamline of the Elettra synchrotron [28]. Immediately after PLD growth, the superlattices were transferred to the ARPES endstation without breaking ultrahigh vacuum to prevent surface contamination. The base pressure of the ARPES chamber was maintained below $5\times10^{-11}$ mbar. Measurements were performed at T = 77 K using a Scienta DA30 hemispherical analyzer operated in deflection mode. The energy resolution was better than 12 meV, and the momentum resolution was 0.018 Å$^{-1}$. The sample was mounted in a normal-emission geometry, with the photon beam incident at 45° relative to the sample surface. Spectra were recorded with linearly polarized light, using both vertical ($E_s$) and horizontal ($E_p$) polarizations. Fermi surface maps and band dispersions were collected at photon energies of 38.5 eV ($\Gamma$ point) and 58 eV (Z point) to probe different $k_z$ planes.

**XMCD Measurements.** X-ray absorption spectroscopy (XAS) and XMCD measurements were conducted at the Magnetic Spectroscopy and Scattering beamline 4.0.2 of the Advanced Light Source (ALS) [30]. The incident photon-energy resolution was approximately 100 meV. Measurements were carried out in the bulk-sensitive luminescence-yield (LY) detection mode at T = 20 K under an applied in-plane magnetic field of 3.5 T.

**Synchrotron-based HAXPES Valence-Band Measurements.** Bulk-sensitive valence-band HAXPES measurements were performed at the P22 beamline [32] of PETRA III at DESY using photon energy of 6.0 keV. At this energy, the inelastic mean free paths of photoelectrons in $CaMnO_3$ and $LaNiO_3$ are estimated to be ~87 and ~71 Å, respectively, corresponding to maximum probing depths of approximately three times these values [33]. The total energy resolution was 380 meV at an analyzer pass energy of 50 eV, and the binding-energy zero was calibrated from the Fermi edge of a standard Au reference. Measurements were carried out at ~77 K.

**Lab-based HAXPES Core-Level Measurements.** Bulk-sensitive HAXPES core-level survey measurements were used to verify the chemical composition of the LNO/CMO superlattices. Measurements were performed using a laboratory-based HAXPES system equipped with a monochromated 5.41 keV X-ray source and a Scienta Omicron EW4000 high-energy hemispherical analyzer. Spectra were acquired at room temperature.

**First-Principles DFT+DMFT Calculations.** First-principles electronic-structure calculations were performed using a combination of density functional theory (DFT) and dynamical mean-field theory (DMFT) [34,35]. DFT calculations were carried out using the Quantum ESPRESSO software package [36,37] within the generalized gradient approximation using the Perdew-Burke-Ernzerhof (PBE) exchange–correlation functional [38]. Ultrasoft pseudopotentials from the GBRV library were employed [39], together with a plane-wave kinetic-energy cutoff of 70 Ry and a charge-density cutoff of 840 Ry. Brillouin-zone integrations were performed using an 7×7×5 Monkhorst-Pack $k$-point mesh, and a Methfessel-Paxton type smearing with width 0.01 Ry was used. Maximally localized Wannier functions were constructed for the low-energy Ni 3$d$ $e_g$ like orbitals at the Fermi level using the Wannier90 formalism [40,41]. The resulting low energy Hamiltonian serves as input for DMFT calculations. Local electronic correlations within the Ni 3$d$ $e_g$ subspace were treated using a rotationally invariant Hubbard-Kanamori interaction, including full spin-flip and pair-hopping terms. The corresponding interaction parameters (U, U′, JH) were chosen as (2.1, 1.5) eV for bulk and (2.1, 1.6) eV for monolayer. The effective quantum impurity model in the DMFT was solved using the numerical exact continuous-time quantum Monte Carlo solver in the hybridization expansion (CT-HYB) as implemented in the TRIQS software library [42,43]. All calculations were performed in the paramagnetic state using TRIQS/solid_dmft [44]. Momentum-resolved spectral functions A(k,ω) were obtained by analytical continuation of the DMFT self-energy performed using Padé approximants. To compare with polarization-dependent ARPES measurements, an orbital dichroism was defined as the difference between the total spectral function and the $d_{x2-y2}$-projected spectral weight.

We consider two structural models that represent the limiting cases of the experimentally studied films: a bulk system, which closely approximates the 6-unit-cell film, and a monolayer (with same k-point density), which serves as a model for the 1-unit-cell film. Both systems are simulated under −1.1% biaxial compressive strain imposed by the CMO substrate, with the lattice vectors constrained to be orthogonal, while the atomic structures retain lower symmetry.

For the monolayer calculations, three pseudocubic buffer layers of LaAlO$_3$ are included to suppress spurious interactions between periodic images. Vacuum is deliberately not used in order to avoid surface effects, and LaAlO$_3$ is chosen as the buffer material due to its wide band gap, which prevents charge transfer to the monolayer. The resulting multilayer heterostructure is nonpolar, as both materials share the same La3+ oxidation state.

**ACKNOWLEDGEMENTS**

The authors gratefully acknowledge Giancarlo Panaccione for valuable discussions and support throughout the project. B.A.C., J.R.P., J.D.G., L.A.L., A.M.D., and A.X.G. acknowledge support from the US Air Force Office of Scientific Research (AFOSR) under award number FA9550-23-1-0476. A.X.G. also gratefully acknowledges the support from the Alexander von Humboldt Foundation. M.K., M.T., and J.C. acknowledge the support by the U.S. Department of Energy, Office of Science, Office of Basic Energy Sciences under award number DE-SC0022160. This work has been partly performed in the framework of the nanoscience foundry and fine analysis (NFFA-MUR Italy Progetti Internazionali) facility. This research also used resources of the Advanced Light Source, which is a DOE Office of Science User Facility under Contract No. DE-AC02-05CH11231. The Flatiron Institute is a division of the Simons Foundation.

**AUTHOR CONTRIBUTIONS**

B.A.C., J.R.P., J.D.G., L.A.L., and A.M.D. performed the spectroscopic experiments under the supervision of A.X.G, and in collaboration with C.K, A.G., C.S., S.K.C., F.M., I.V., and P.O. All theoretical calculations were carried out by A.H. and S.B. The samples were synthesized and characterized by S.K.C., F.M., and P.O. in collaboration with M.T., M.K., and J.C. Additional theoretical support was provided by A.J.M. All authors contributed to the writing of the manuscript.

# Supplementary Information

# Dimensionality-Driven Electronic and Orbital Transitions Mediating Interfacial Magnetism in LaNiO$_3$/CaMnO$_3$ Observed *In Situ*

B-A. Courchene[1], A. Hampel[2], S. Beck[2], J. R. Paudel[1,3], J. D. Grassi[1,4], L. A. Lapinski[1], A. M. Derrico[1,5], M. Terilli[6], M. Kareev[6], C. Klewe[7], A. Gloskovskii[8], C. Schlueter[8], S. K. Chaluvadi[9], F. Mazzola[9,10], I. Vobornik[9], P. Orgiani[9,11], J. Chakhalian[2], A. J. Millis[2,12], and A. X. Gray[1,*]

[1] *Department of Physics, Temple University, Philadelphia, Pennsylvania 19122, USA*
[2] *Center for Computational Quantum Physics, Flatiron Institute, New York, New York 10010, USA*
[3] *Chemical Sciences Division, Lawrence Berkeley National Laboratory, Berkeley, California 94720, USA*
[4] *Department of Physics, Cornell University, Ithaca, New York 14853, USA*
[5] *Department of Physics, University of California, Berkeley, California 94720, USA*
[6] *Department of Physics and Astronomy, Rutgers University, Piscataway, New Jersey 08854, USA*
[7] *Advanced Light Source, Lawrence Berkeley National Laboratory, Berkeley, California 94720, USA*
[8] *Deutsches Elektronen-Synchrotron, DESY, 22607 Hamburg, Germany*
[9] *CNR-IOM Istituto Officina dei Materiali, Trieste 34149, Italy*
[10] *Dipartimento di Fisica e Astronomia, Università degli Studi di Padova, Padua 35131, Italy*
[11] *AREA Science Park, Trieste 34149, Italy*
[12] *Department of Physics, Columbia University, New York, New York 10027, USA*
*email: axgray@temple.edu

## 1. *In situ* Low-Energy Electron Diffraction (LEED) and Scanning Tunneling Microscopy (STM) Characterization

LEED patterns acquired *in situ* immediately after growth confirmed the epitaxial order and high surface crystallinity of the topmost LaNiO$_3$ (LNO) layer in both the single-layer LNO sample (Fig. S1a) and the LNO/CMO superlattice (Fig. S1b). Figure S1a shows the LEED diffraction pattern of a single LNO film deposited on a LaAlO$_3$ (001) substrate, acquired with an incident electron energy of 135 eV, and reveals a well-defined surface periodicity indicative of high crystalline order and epitaxial growth. Figure S1b shows the LEED diffraction pattern of the 4CMO/6LNO superlattice, recorded at 70 eV, demonstrating an equally well-ordered surface structure and preserved epitaxial registry of the superlattice. For the same LaNiO₃ thin film, *in situ* STM images (Figs. S1c,d) show a flat surface morphology with an RMS roughness of approximately 0.3 nm, corresponding to less than one unit cell.

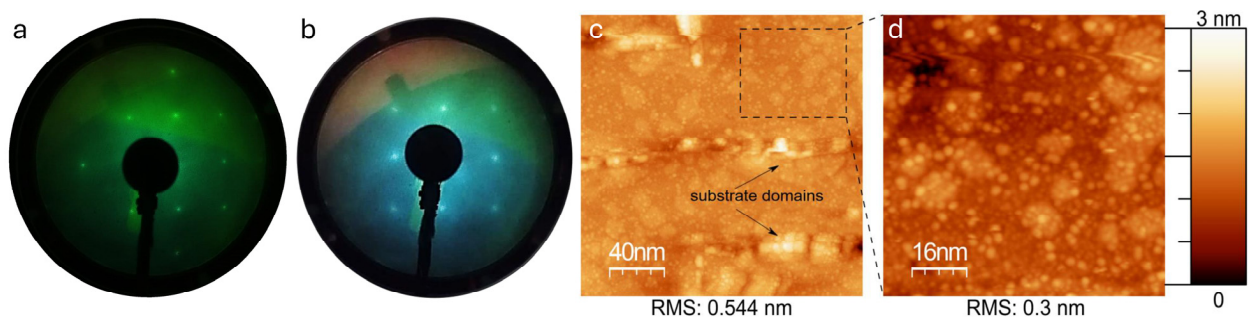

**Figure S1 | a,b,** *In-situ* LEED patterns measured immediately after growth from (a) a LaNiO₃ thin film on LaAlO₃(001) and (b) the 4CMO/6LNO superlattice. In both cases, the sharp diffraction spots are consistent with high surface crystallinity and epitaxial order. The incident electron energies were 135 eV for a and 70 eV for b. **c,d,** *In-situ* STM images of the same LaNiO₃ film, showing a smooth surface with an RMS roughness of about 0.3 nm, below one unit cell. The LaAlO₃(001) substrate in panel (c) shows typical structural twin-domain-related surface corrugations.

## 2. *Ex situ* X-ray Diffraction (XRD) and X-ray Reflectivity (XRR) Characterization

Figures S2a-d show the *ex situ* laboratory-based XRD $\theta$-$2\theta$ spectra measured for all four superlattices. The LaAlO₃ (002) substrate peak appears at $2\theta=48^0$, consistent with prior studies. The first-order superlattice peaks (SL₋₁) and the superlattice thickness fringes exhibit line shapes characteristic of high-quality single-crystalline superlattices. Figures S2e-h show the complementary laboratory-based angle-resolved XRR spectra measured on the same superlattices (blue curves). The pronounced thickness fringes indicate high-quality interfaces. Best fits to the experimental spectra (red curves), obtained using self-consistent X-ray optical modeling, yield individual LNO and CMO layer thicknesses in good agreement with the expected values.

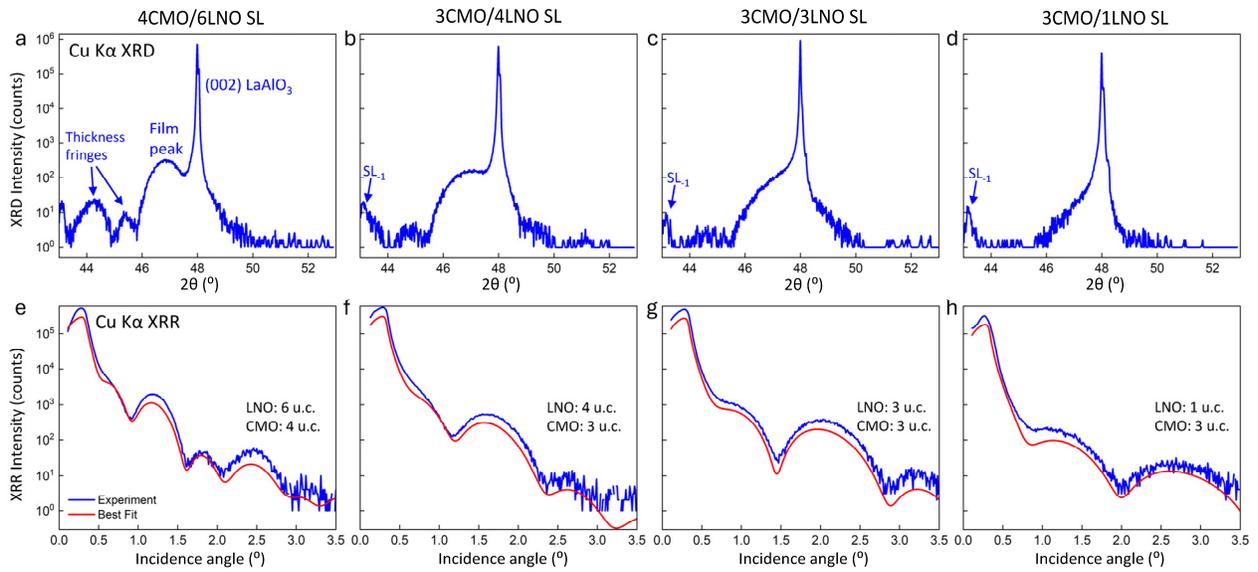

**Figure S2 | a-d**, Laboratory-based XRD θ-2θ scans for the four superlattices studied here: (a) 4 u.c. $CaMnO_3$ / 6 u.c. $LaNiO_3$, (b) 3 u.c. $CaMnO_3$ / 4 u.c. $LaNiO_3$, (c) 3 u.c. $CaMnO_3$ / 3 u.c. $LaNiO_3$, and (d) 3 u.c. $CaMnO_3$ / 1 u.c. $LaNiO_3$. **e-h**, Corresponding XRR spectra from the same samples (blue curves), shown in the same order as in a-d. The well-defined thickness fringes are consistent with smooth, high-quality interfaces. The red curves show fits obtained from self-consistent X-ray optical modeling, from which the individual LNO and CMO layer thicknesses were determined and found to be in good agreement with the nominal values.

### 3. Hard X-ray Photoelectron Spectroscopy (HAXPES) Chemical Characterization

The nominal chemical composition of the superlattices was confirmed by bulk-sensitive HAXPES measurements performed with a laboratory-based spectrometer equipped with a monochromated 5.41 keV X-ray source and a Scienta Omicron EW4000 high-energy hemispherical analyzer. Figure S3 shows wide-energy-range HAXPES survey spectra for all four superlattices. The presence of all expected elements - Ca, Mn, O, La, and Ni, as well as C from the surface-adsorbed C/O contamination layer, is confirmed by the corresponding core-level peaks. In addition, the relative intensities of the Ni $2s$ and Mn $2p$ core-level peaks follow the expected trend as the thickness of the top $LaNiO_3$ layer increases from 1 u.c. to 6 u.c., with the Ni $2s$ intensity increasing and the Mn $2p$ intensity decreasing accordingly.

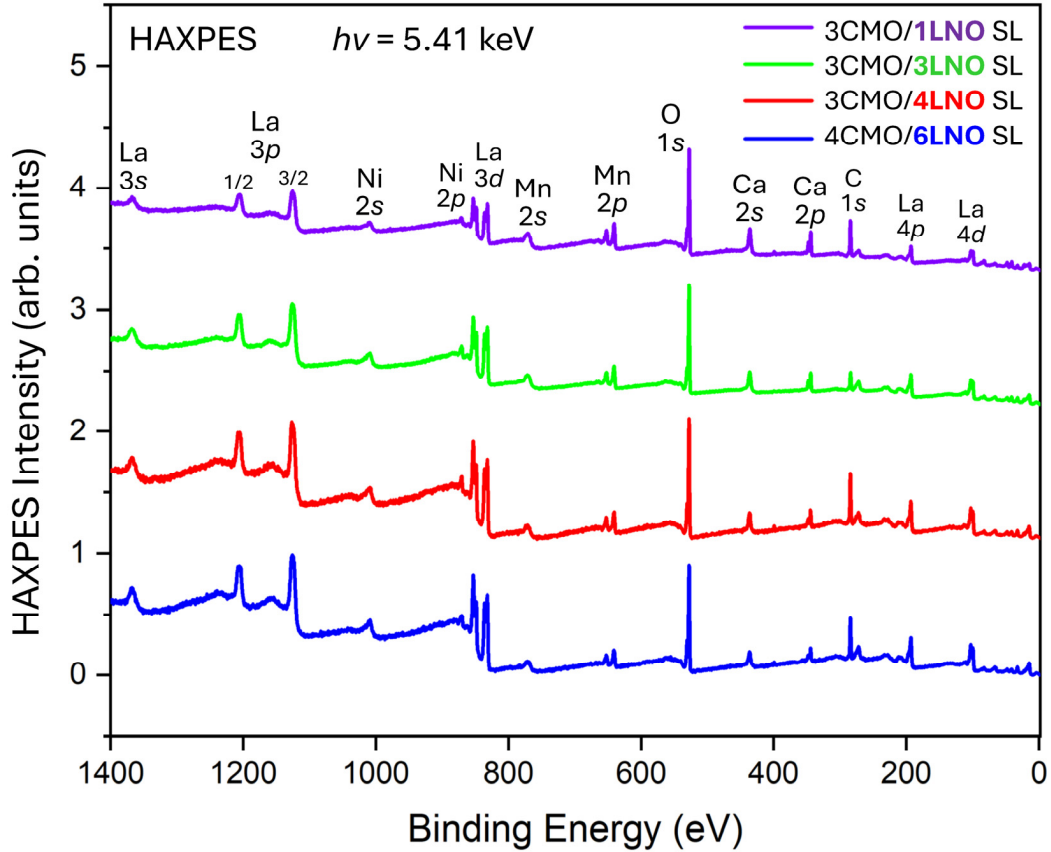

**Figure S3** | Wide-energy-range HAXPES survey spectra of the four superlattices, confirming the expected elemental composition and the systematic evolution of the Ni 2s and Mn 2p peak intensities with increasing top LaNiO3 layer thickness. Measurements were carried out at room temperature.

## 4. Hard X-ray Photoelectron Spectroscopy (HAXPES) Valence-Band Characterization

Angle-integrated valence-band HAXPES spectra for all four superlattices, at the P22 beamline of PETRA III at DESY using photon energy of 6.0 keV, reveal a systematic suppression of the near-$E_F$ spectral weight from the strongly hybridized Ni 3d $e_g$ and $t_{2g}$ states with decreasing LaNiO3 thickness (see Figure S4). This bulk-sensitive evolution signals the onset of the metal-insulator transition in 3CMO/3LNO SL and a fully developed gap in 3CMO/1LNO SL.

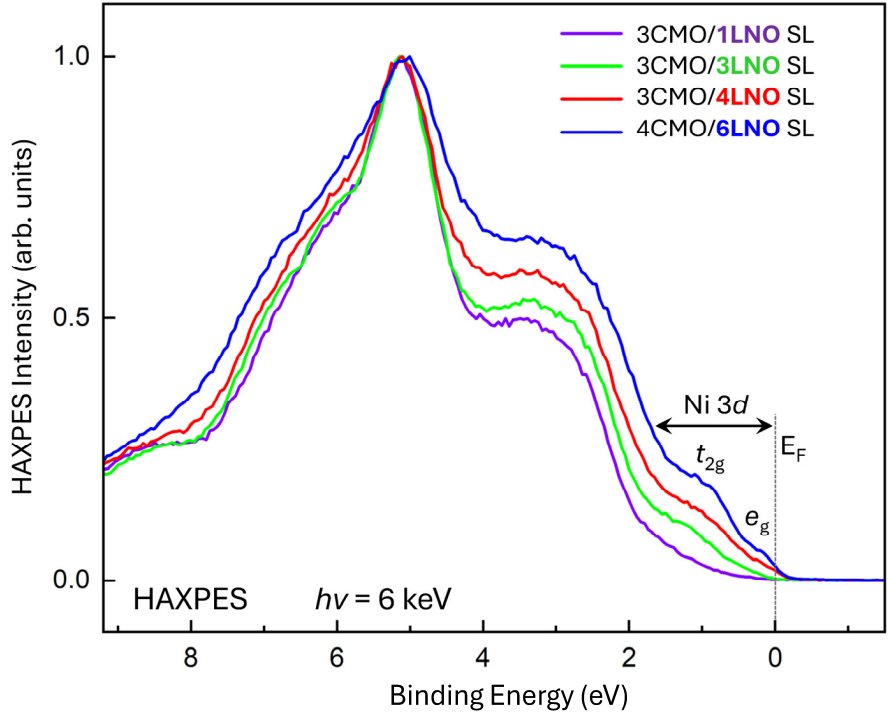

**Figure S4** | Angle-integrated valence-band HAXPES spectra for all four superlattices, acquired at a photon energy of 6.0 keV with the total energy resolution of 380 meV (at a pass energy of 50 eV).

## 5. Momentum-Integrated, Orbital-Resolved DFT+DMFT Spectral Functions of Bulk and Monolayer LaNiO$_3$

Figure S5 on the next page presents the momentum-integrated, orbital-resolved DFT+DMFT spectral functions calculated for bulk and monolayer LaNiO$_3$. In the bulk case (a), the total spectral function $A_{tot}(\omega)$ is shown together with its $d_{x^2-y^2}$ and $d_{z^2}$ -resolved components, allowing direct comparison of the orbital character across the valence and near-Fermi-level energy range. The lower panel displays the dichroic signal $\Delta A(\omega)$, defined as the difference between the total spectral function $A_{tot}(\omega)$ and the $d_{x^2-y^2}$ -projected contribution $A_{dx^2-y^2}(\omega)$. The same analysis is shown for the monolayer in Fig. S5b. Comparison of the two cases highlights the change in orbital-resolved spectral weight upon reducing the dimensionality from bulk to the monolayer limit.

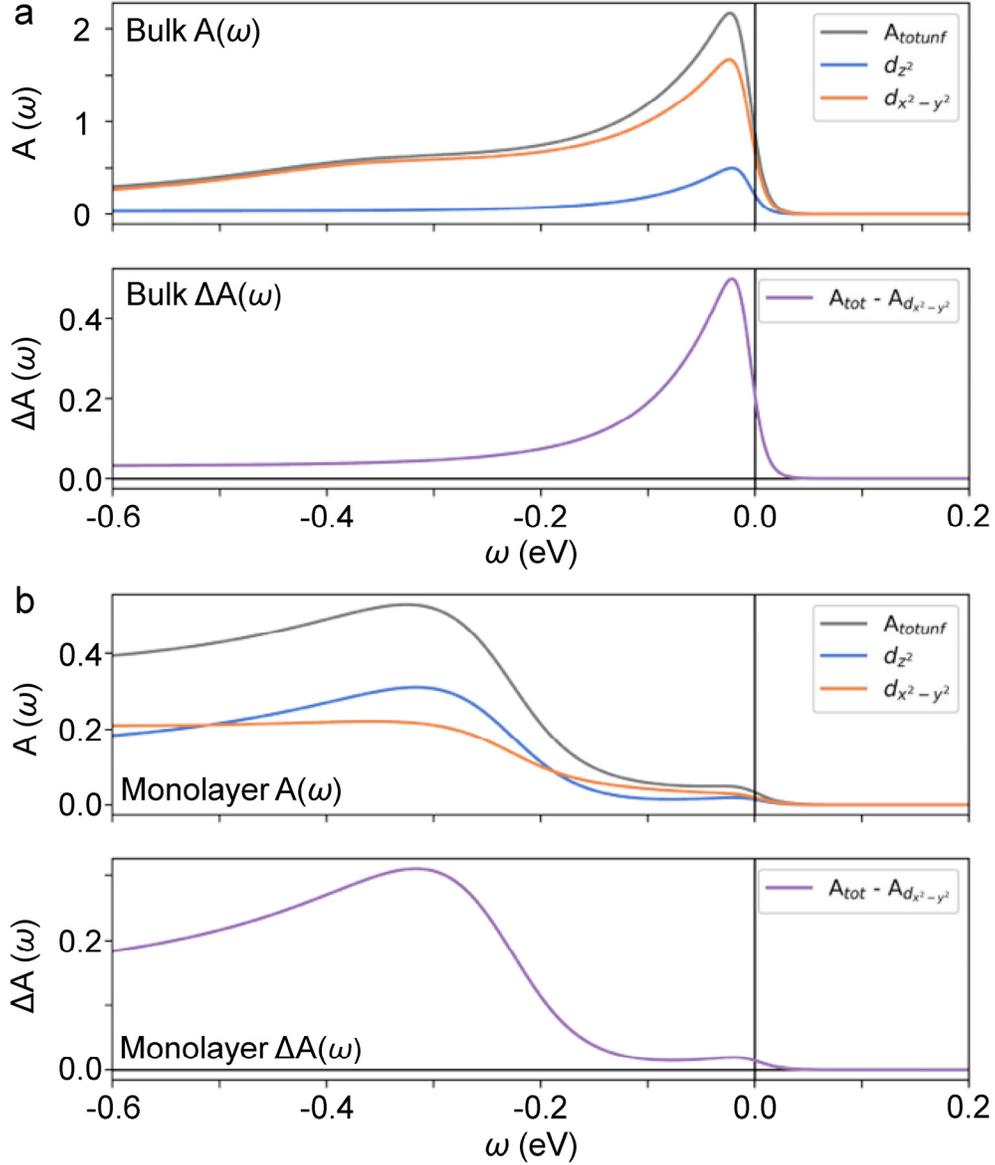

**Figure S5 | a,** Momentum-integrated, orbital-resolved DFT+DMFT spectral functions for bulk LaNiO$_3$. The gray curve denotes the total spectral function, $A_{tot}(\omega)$, and the orange and blue curves denote the $d_{x2-y2}$ and $d_{z2}$ projected components, respectively. The purple curve in the lower panel of a shows the dichroic signal, defined as $\Delta A(\omega) = A_{tot}(\omega) - A_{dx2-y2}(\omega)$. b. Same as (a), but for monolayer LaNiO$_3$.